\documentclass[preprint]{aastex}

\shorttitle{The Globular Cluster System}
\shortauthors{van den Bergh}

\begin{document}

\title{Some Global Characteristics of the Galactic Globular Cluster System}

\author{Sidney van den Bergh}
\affil{Dominion~Astrophysical~Observatory, Herzberg~Institute~of~Astrophysics, National~Research~Council~of~Canada, 5071~W.~Saanich~Rd., Victoria,~British~Columbia, V9E~2E7, Canada}
\email{sidney.vandenbergh@nrc-cnrc.gc.ca}

\begin{abstract}
The relations between the luminosities $M_{V}$, the metallicities $[Fe/H]$, the Galactocentric radii $R$, and the central concentration indices $c$ of Galactic globular clusters are discussed. It is found that the most luminous clusters rarely have collapsed cores. The reason for this might be that the core collapse time scales for such populous clusters are greater than the age of the Galaxy. Among those clusters, for which the structure has not been modified by core collapse, there is a correlation between central concentration and integrated luminosity, in the sense that the most luminous clusters have the strongest central concentration. The outermost region of the Galaxy with $R>10$ kpc was apparently not able to form metal-rich $([Fe/H]>-1.0)$ globular clusters, whereas such clusters (of which Ter 7 is the prototype) were able to form in some nearby dwarf spheroidal galaxies. It is not yet clear how the popular hypothesis that globular clusters were initially formed with a single power law mass spectrum can be reconciled with the observation that both (1) Galactic globular clusters with $R>80$ kpc, and (2) the globulars associated with the Sagittarius dwarf, appear to have bi-modal luminosity functions.\end{abstract}

\keywords{globular clusters: general}

\section{INTRODUCTION}

Globular clusters are among the oldest objects to have formed in the Galaxy. They therefore provide valuable information on the early evolutionary history of the Milky Way System. The present paper discuses the interrelationships between integrated cluster magnitude $M_{V}$, cluster metallicity $[Fe/H]$, 
Galactocentric distance $R$, and the central concentration index $c=\log (r_{t}/r_{c}) $\citep{kin62}. The basic observational data, which are listed in Table 1, were drawn from \citet{har96}, but updated from new information that is compiled at http://physun.physics.mcmaster.ca/Globular.html. A discussion, in some ways similar to the present one (but based on older and less complete data), was previously published by \citet{djo94}. That paper also 
gives extensive references to earlier work on this subject. Traditionally \citep{kin59,zin85} the Galactic globular cluster system is regarded as consisting of a metal-rich disk and a metal-poor halo.  However, more recently \citep{min95,bar99,cot99,vdb00a} it has become clear that the metal-rich clusters actually constitute a bulge, rather than a disk, population. Among the questions to which answers will be sought in the present investigation are the following: (1) Is the Galactic globular cluster system old enough for even the most massive globular clusters to have collapsed? (2) Does the structure of pre-collapse clusters depend on their mass or environment? (3) Do metal-poor globular clusters with $[Fe/H]<-1.0$ exhibit a radial metallicity gradient over the range $-0.5<\log R($kpc$) < 2.5$? (4) Do metal-rich clusters with $[Fe/H]>-1.0$ show a radial abundance gradient over the range $-0.5<\log R<1.0$? (5) Did the  metal-rich globular clusters in the outer reaches of the Galaxy have a different origin from the metal-rich globulars with $R<10$ kpc that are embedded in the main body of the Galaxy and its inner halo? Questions (4) and (5) above may be rephrased by asking if the Galaxy contains physically distinct metal-rich and metal-poor cluster populations. The presence, or absence, of metallicity gradients within these two populations might throw some light on the evolutionary history of each of these two populations. Finally it is noted that attempts to answer some of these questions are intrinsically uncertain because of the relatively small number of globular clusters associated with the Galaxy.

\section{CORRELATIONS}

All six possible permutations of the relations between $M_{V}$, $[Fe/H]$, $R$ and $c$ are plotted in Figure 1. A brief discussion of observational data on these relations is given below.

\subsection{$[Fe/H]$ {\it versus} $\log R$}

This plot shows the well know dichotomy between the metal-poor clusters with $<[Fe/H]> \approx \space-1.5$ and the metal-rich clusters with $<[Fe/H]>$\space $\approx -0.5$. All but three of the metal-rich $([Fe/H]>-1.0)$ globulars are located at $R<10$ kpc, i.e., almost all metal-rich clusters are located in the inner region of the Galaxy. Among these exceptions two (Terzan 7 and Palomar 12) seem to be associated with the tidally disintegrating Sagittarius dwarf galaxy \citep{iba94,irw94}. The third exception is the cluster Palomar 1, which appears to be a ``young'' globular \citep{ric96} that might have formed in a no longer extant dwarf spheroidal \citep{vdb00b}. In summary, 43 out of 46 (93\%) of metal-rich Galactic globular clusters with $[Fe/H]>-1.0$ are old objects located at $R<10$ kpc. The three remaining clusters are thought to be younger objects that formed (or may have formed) in dwarf spheroidal galaxies.
   Is there a radial metallicity gradient among metal-poor globular clusters? Such a gradient would be expected \citep{egg62} if the young Galactic halo was able to enrich itself in heavy elements while it was contracting. On the other hand, such a gradient would not be expected for a more chaotic assembly of the halo, such as has been proposed by \citet{sea78}. Data on the radial distributions of moderately metal-poor $(-1.60<[Fe/H]\leq -1.00)$ and very metal-poor $(-2.40<[Fe/H]\leq -1.60)$ clusters are collected in Table 2. Separate information is provided for (1) all metal-poor clusters, and (2) with probable companions of the Sagittarius dwarf excluded. Inspection of these data hints at a possible metallicity gradient in the expected sense, i.e., with a small excess of the highest metallicity subgroup at small Galactocentric radii. However, a Kolmogorov-Smirnov test shows only an 88\% probability that this effect is real for the entire sample, and an 82\% probability that it is real for the sample from which probable companions to the Sagittarius galaxy have been excluded. In summary, it is concluded that the population of metal-poor Galactic halo clusters is too small to establish with certainty if they exhibit a metallicity gradient. For metal-rich globular clusters with $[Fe/H]>-1.0$ there is no statistically significant evidence for a correlation between $[Fe/H]$ and $R$.
  Finally, it is noted that we presently do not understand how some nearby dwarf spheroidal galaxies were able to form quite metal-rich globular $([Fe/H]<-1.0)$ clusters. This contrasts with the situation in the outer $(R>10$ kpc$)$ regions of the Galaxy that were apparently never able to form similarly metal-rich globular clusters. This result suggests that the specific globular cluster frequency in the Galaxy may have exhibited a steep decline with increasing metallicity, whereas the specific cluster frequency may have been less sensitive to metallicity in (some) dwarf spheroidal galaxies.

\subsection{$[Fe/H]$ {\it versus} $M_{V}$}

The metallicity distribution of Galactic globular clusters does not appear to depend on luminosity. Kolmogorov-Smirnov tests show no significant differences between the metallicity distributions of objects brighter and fainter than $M_{V}=-5.0$. The same conclusion holds if the sample is divided at $M_{V}=-6.0$. This result is somewhat surprising because one might have expected faint metal-rich clusters (which mainly occur in high density regions) to have been preferentially destroyed by bulge and disk shocks, or eroded by tidal stripping. On the other hand the most massive (luminous) clusters are expected to be survivors e.g., \citet{gne02}.

\subsection{$[Fe/H]$ {\it versus} $c$}

The Figure shows little (or no) evidence for a dependence of metallicity on the central concentration of index $c$ \citep{kin62} of Galactic globular clusters. In particular the fraction of globulars with collapsed cores ($c\approx 2.5$) does not appear to depend strongly on metallicity.

\subsection{$c$  {\it versus} $\log R$}

The overwhelming majority of clusters with collapsed cores, i.e., those with $c\approx 2.5$, are found to be located at $R<10$ kpc. The reason for this is, no doubt, that globular cluster half-light radii tend to decrease with decreasing Galactocentric distance \citep{vdb84}. As a result the typical relaxation times (and core collapse time scales) of clusters near the Galactic center are significantly shorter than they are for globulars at larger Galactocentric distances. Among metal-rich $([Fe/H]>-1.0)$ clusters the central concentration index $c$ correlates with Galactocentric distance in the sense that the most metal-rich clusters have the strongest central concentration of light. For metal-rich clusters with $c<1.00,<\log R>$ \space $=0.69 \pm 0.07$, which is significantly larger than $<\log R>$ \space $=0.34 \pm 0.10$ for the metal-rich clusters with $c>2.00$. A similar correlation for metal-poor clusters appears to exhibit more scatter than does that among the metal-rich Galactic globulars. The observation that clusters with low central concentration, on average, have larger Galactocentric distances than do those at larger distances is (see Section 2.5) due to the fact that most faint clusters with $M_{V}>-6.0$ have low central concentrations ($c<1.0$), whereas the majority of clusters $1.5<c<2.4$ are more luminous than $M_{V}=-6.0$.

\subsection{$c$ {\it versus} $M_{V}$}

The structure of clusters may be characterized by their central concentration index $c$. A plot of $c$ versus $M_{V}$ exhibits a clear dichotomy between clusters that have collapsed cores with $c\approx 2.5$, and those with smaller $c$ values that do not exhibit such collapsed cores. It is of interest to note that only one Galactic globular cluster with $M_{V}<-8.0$ is known to have a collapsed core. This is NGC 7078. The reason for the observed paucity of luminous clusters with collapsed cores ($c=2.5$) might be that the core collapses time scale, which scales as $N / ln N$, is too long for the most massive clusters to experience core collapse within a Hubble time. However, a possible argument against this view is that few of the clusters for which \citet{har96} lists half-mass relaxtion times have relaxation times longer than 5 Gyr. A comparison between the frequency distributions of clusters with collapsed cores ($c>2.4$), and clusters of intermediate central concentration ($1.5<c<2.0$), is shown in Table 3. A Kolmogorov-Smirnov test shows that there is a 99.1\% probability that the intermediate concentration clusters, and the collapsed core clusters, were not drawn from the same parent luminosity distribution. A plot of $c$ versus $M_{V}$, for those clusters that do not have collapsed cores, shows a close correlation between central concentration and luminosity. [The exact value of the correlation coefficient is sensitive to the choice of clusters that are omitted from the calculation because they are regarded as objects that are on their way to core collapse.] Among such clusters lacking collapsed cores the most concentrated objects have the highest luminosities. This correlation was previously also noted by \citet{djo94}. Clusters on the ``main sequence'' that lack collapsed cores have central concentrations that range from $c\approx 0.6$ at $M_{V}\approx -5$, to $c\approx 1.8$ at $M_{V}\approx -9$. It seems likely that this observed correlation between luminosity and central concentration (of clusters that do not have collapsed cores) is due to initial conditions at the time of cluster formation. A few clusters, such as NGC6397 and NGC6717 lie between the ``main sequence'' and the sequence of objects with collapse cores at $c\approx 2.5$. Possibly such clusters are presently moving toward, but have not yet attained, core collapse. In view of the rich diversity of the relation between $c$ and $M_{V}$, that is shown in Figure 1, it is remarkable \citep{vdb91} that the half-light radii of Galactic globular clusters turns out to be independent of cluster luminosity. It is noted in passing that most of the globulars associated with the Sagittarius system appear to lie on (or close to) the ``main sequence'' for Galactic globular clusters in the $M_{V}$ versus $c$ diagram.

\subsection{$\log R$ {\it versus} $M_{V}$}

In the $M_{V}$ versus $\log R$ plane centrally concentrated clusters with $c>2.0$ are all found to lie below the line
\begin{equation}
\log R=-0.43-0.21M_{V}  
\end{equation}

i.e., luminous compact clusters mainly occur at small values of $\log R$, whereas faint extended clusters mostly lie at large Galactocentric radii.  Inspection of the $M_{V}$ versus $\log R$ diagram shows a clear distinction between the luminosity distributions of globular clusters in the main body of the Galaxy and those at large Galactocentric distances. Globular clusters with $R<80$ kpc exhibit a near Gaussian\footnote[1]{The ``tail'' of faint clusters with $M_{V}>-4.0$ may represent objects that were initially more luminous, but that were subsequently decimated by disk (or core) shocks and subsequent tidal detachment of stars. This hypothesis is supported by the observation that the faint cluster E3 $(M_{V}=-2.8$, $R=7.6$ kpc) is unusually rich in binary stars (van den Bergh 1980).} luminosity function that peaks at $M_{V}\approx -7.5$, while the outer halo clusters have either $M_{V}\approx -5$ or $M_{V}\approx -10$. Surprisingly (van den Bergh 2000a, p. 228) the luminosity function of the globular clusters associated with the Sagittarius dwarf system also has a bi-modal luminosity function that appears to resemble that of the clusters with $R>80$ kpc in the outer halo of the Galaxy. [If one includes the clusters that may be associated with the Sagittarius, then the luminosity function might perhaps be better described as being weighted towards objects of very low luminosity (mass)]. Van den Bergh (2000a, p. 229) has used a Kolmogorov-Smirnov test to show that there is only a 4\% probability that the luminosity distribution of Galactic globulars with $R<80$ kpc were drawn from the same parent population as those as those with $R>80$ kpc.
   It is widely believed \citep{fal77} that the presently observed globular clusters are the survivors of an initially much larger population. Within the main body of the Galaxy the present luminosity function of globular clusters has a near Gaussian shape \citep{abr95}. It is thought that this is due to the dynamical ``erosion'' \citep{ost72,gne02} of what was initially a power law mass spectrum. However, it is difficult to see how such a process could could produce the apparently bi-modal luminosity distribution of (1) the Galactic globular clusters at $R>80$ kpc, and (2) of the globular clusters that appear to be associated with the Sagittarius dwarf. Perhaps outer halo clusters such as N2419 ($M_{V}=-9.6$) and Pal.3 ($M_{V}=-5.7$) actually belong to separate populations with very different evolutionary histories. By the same token the cluster N6715 (=M 54) with $M_{V}=-10.0$, which is located at the center of the Sagittarius dwarf, may have had a different evolutionary history than that of the (and non-central) Sagittarius companions Ter 7, Ter 8, and Arp 2, all of which are faint and have $M_{V}\approx -5$.

\section{CONCLUSIONS}

The following conclusions may be drawn from the presently available data on Galactic globular clusters:
\begin{itemize}
\item Collapsed cores are rare among the most luminous Galactic globulars. This might suggest that the the collapse time scale for the most massive globular clusters is longer than the age of the Galaxy.
\item Most of the globular clusters that do not have collapsed cores show a strong correlation between luminosity and central concentration. This correlation is in the sense that the most luminous clusters have the strongest central concentration of light. 
\item Metal-poor clusters with $[Fe/H]<-1.0$ occur at all Galactocentric distances, whereas all but three of the metal-rich clusters are situated at $R<10$ kpc.
\item These three metal-rich ($[Fe/H]>-1.0$) clusters at $R>10$ kpc may originally have formed in, or in association with, dwarf spheroidal companions to the Galaxy.
\item For metal-rich clusters with $[Fe/H]>-1.0$ there is no evidence for the existence of a radial metallicity gradient. The data hint at, but are not numerous enough to establish, the reality of a radial metallicity gradient among metal-poor clusters with $[Fe/H]<-1.0$.
\item Both the outer halo globular clusters with $R>80$ kpc, and those associated with the Sagittarius dwarf spheroidal, appear to have bi-modal luminosity functions. It is difficult to see how such bi-modal luminosity distributions could have arisen from clusters that initially had a single power law mass spectrum.
\end{itemize}

   The luminosity $M_{V}$, the metallicity $[Fe/H]$, the Galactocentric distance $R$, and the concentration parameter $c$, are presently known for the great majority of Galactic globular clusters. Nevertheless, the relatively small total number of Galactic globular clusters places significant statistical restrictions on the degree of certainty with which correlations between many of these parameters can be established. In particular the Galactic globular cluster population is too small to establish if there exists a radial metallicity gradient among the metal-poor $([Fe/H]<-1.0)$ halo cluster population. By the same token it is not possible to say if the metal-rich bulge/disk globular clusters with $[Fe/H]>-1.0$ have a radial composition gradient.

\acknowledgments

   It is a pleasure to thank Russell Redman for his kind help with Figure 1. I also thank Scott Tremaine and the referee for helpful comments.

\clearpage

\begin{figure}
\plotone{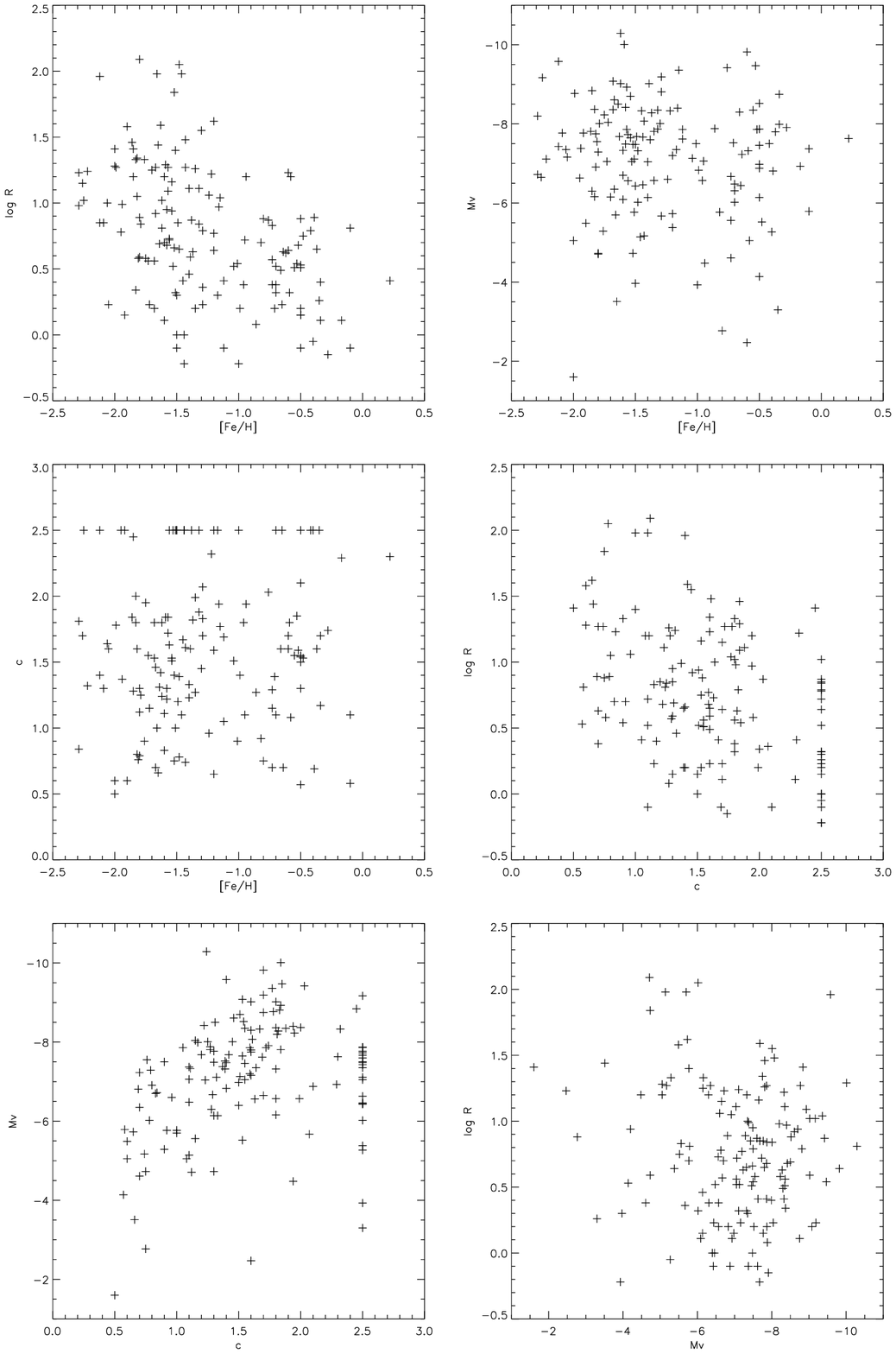}
\caption{Plots of all possible combinations of the metallicity $[Fe/H]$, the central concentration index $c$, the visual absolute magnitude $M_{V}$ and the logarithm of the Galactocentric distance $R$, for Galactic globular clusters. 
\label{fig1}}
\end{figure}

\clearpage

\begin{deluxetable}{lrrrr}
\tabletypesize{}
\tablecaption{Galactic globular cluster parameters\label{tbl-1}}
\tablewidth{0pt}
\tablehead{\colhead{Cluster} & \colhead{$[Fe/H]$} & \colhead{$c$} & \colhead{$M_V$} & \colhead{$\log R$}}
\startdata

N 104    &-0.76    &2.03   &-9.42    &0.87\\
N 288    &-1.24    &0.96   &-6.60    &1.06\\
N 362    &-1.16    &1.94   &-8.40    &0.97\\
N1261    &-1.35    &1.27   &-7.81    &1.26\\
Pal 1    &-0.60    &1.60   &-2.47    &1.23\\
AM 1     &-1.80    &1.12   &-4.71    &2.09\\
Eri      &-1.46    &1.10   &-5.14    &1.98\\
Pal 2    &-1.30    &1.45   &-8.01    &1.55\\
N1851    &-1.22    &2.32   &-8.33    &1.22\\
N1904    &-1.57    &1.72   &-7.86    &1.27\\
N2298    &-1.85    &1.28   &-6.30    &1.20\\
N2419    &-2.12    &1.40   &-9.58    &1.96\\
Pyx      &-1.20    &0.65   &-5.73    &1.62\\
N2808    &-1.15    &1.77   &-9.36    &1.04\\
E3       &-0.80    &0.75   &-2.77    &0.88\\
Pal 3    &-1.66    &1.00   &-5.70    &1.98\\
N3201    &-1.58    &1.30   &-7.49    &0.95\\
Pal 4    &-1.48    &0.78   &-6.02    &2.05\\
N4147    &-1.83    &1.80   &-6.16    &1.33\\
N4372    &-2.09    &1.30   &-7.77    &0.85\\
Rup 106  &-1.67    &0.70   &-6.35    &1.27\\
N4590    &-2.06    &1.64   &-7.35    &1.00\\
N4833    &-1.79    &1.25   &-8.01    &0.84\\
N5024    &-1.99    &1.78   &-8.77    &1.27\\
N5053    &-2.29    &0.84   &-6.72    &1.23\\
N5139    &-1.62    &1.24  &-10.29    &0.81\\
N5272    &-1.57    &1.84   &-8.93    &1.09\\
N5286    &-1.67    &1.46   &-8.61    &0.92\\
AM 4     &-2.00    &0.50   &-1.60    &1.41\\
N5466    &-2.22    &1.32   &-7.11    &1.24\\
N5634    &-1.82    &1.60   &-7.75    &1.34\\
N5694    &-1.86    &1.84   &-7.81    &1.46\\
I4499    &-1.60    &1.11   &-7.33    &1.20\\
N5824    &-1.85    &2.45   &-8.84    &1.41\\
Pal 5    &-1.43    &0.74   &-5.17    &1.27\\
N5897    &-1.80    &0.79   &-7.29    &0.89\\
N5904    &-1.29    &1.83   &-8.81    &0.79\\
N5927    &-0.37    &1.60   &-7.80    &0.65\\
N5946    &-1.38    &2.50   &-7.60    &0.87\\
BH 176 &\nodata &\nodata   &-4.20    &0.94\\
N5986    &-1.58    &1.22   &-8.42    &0.68\\
Lyn 7    &-0.62 &\nodata &\nodata    &0.62\\
Pal 14   &-1.52    &0.75   &-4.73    &1.84\\
N6093    &-1.75    &1.95   &-8.23    &0.58\\
N6121    &-1.20    &1.59   &-7.20    &0.77\\
N6101    &-1.82    &0.80   &-6.91    &1.05\\
N6144    &-1.73    &1.55   &-7.05    &0.56\\
N6139    &-1.68    &1.80   &-8.36    &0.56\\
Ter 3    &-0.73    &0.70   &-4.61    &0.38\\
N6171    &-1.04    &1.51   &-7.13    &0.52\\
1636-    &-1.50 &\nodata   &-3.97    &0.30\\
N6205    &-1.54    &1.51   &-8.70    &0.94\\
N6229    &-1.43    &1.61   &-8.07    &1.48\\
N6218    &-1.48    &1.39   &-7.32    &0.65\\
N6235    &-1.40    &1.33   &-6.14    &0.46\\
N6254    &-1.52    &1.40   &-7.48    &0.66\\
N6256    &-0.70    &2.50   &-6.02    &0.32\\
Pal 15   &-1.90    &0.60   &-5.49    &1.58\\
N6266    &-1.29    &1.70   &-9.19    &0.23\\
N6273    &-1.68    &1.53   &-9.08    &0.20\\
N6284    &-1.32    &2.50   &-7.87    &0.84\\
N6287    &-2.05    &1.60   &-7.16    &0.23\\
N6293    &-1.92    &2.50   &-7.77    &0.15\\
N6304    &-0.59    &1.80   &-7.32    &0.32\\
N6316    &-0.55    &1.55   &-8.35    &0.51\\
N6341    &-2.29    &1.81   &-8.20    &0.98\\
N6325    &-1.17    &2.50   &-7.35    &0.30\\
N6333    &-1.72    &1.15   &-8.04    &0.23\\
N6342    &-0.65    &2.50   &-6.44    &0.23\\
N6356    &-0.50    &1.54   &-8.52    &0.88\\
N6355    &-1.50    &2.50   &-7.48    &0.00\\
N6352    &-0.70    &1.10   &-6.48    &0.52\\
I1257    &-1.70 &\nodata   &-6.15    &1.25\\
Ter 2    &-0.40    &2.50   &-5.27   &-0.05\\
N6366    &-0.82    &0.92   &-5.77    &0.70\\
Ter 4    &-1.60 &\nodata   &-6.09    &0.11\\
HP 1     &-1.50    &2.50   &-6.43   &-0.10\\
N6362    &-0.95    &1.10   &-7.06    &0.72\\
Lil 1    &+0.22    &2.30   &-7.63    &0.41\\
N6380    &-0.50    &1.55   &-7.46    &0.51\\
Ter 1    &-0.35    &2.50   &-3.30    &0.26\\
Ton 2    &-0.50    &1.30   &-6.14    &0.15\\
N6388    &-0.60    &1.70   &-9.82    &0.64\\
N6402    &-1.39    &1.60   &-9.02    &0.59\\
N6401    &-1.12    &1.69   &-7.62   &-0.10\\
N6397    &-1.95    &2.50   &-6.63    &0.78\\
Pal 6    &-0.10    &1.10   &-7.37   &-0.10\\
Dj 1   &\nodata    &1.50   &-6.40    &0.00\\
Ter 5    &-0.28    &1.74   &-7.91   &-0.15\\
N6440    &-0.34    &1.70   &-8.75    &0.11\\
N6441    &-0.53    &1.85   &-9.47    &0.54\\
Ter 6    &-0.50    &2.50   &-7.87    &0.20\\
N6453    &-1.53    &2.50   &-7.05    &0.52\\
UKS 1    &-0.50    &2.10   &-6.88   &-0.10\\
N6496    &-0.64    &0.70   &-7.23    &0.63\\
Ter 9    &-1.00    &2.50   &-3.93   &-0.22\\
Dj 2     &-0.50    &1.50   &-6.98    &0.15\\
N6517    &-1.37    &1.82   &-8.28    &0.63\\
Ter 10   &-0.70 &\nodata   &-6.31    &0.38\\
N6522    &-1.44    &2.50   &-7.67   &-0.22\\
N6535    &-1.80    &1.30   &-4.73    &0.59\\
N6528    &-0.17    &2.29   &-6.93    &0.11\\
N6539    &-0.66    &1.60   &-8.30    &0.49\\
N6540    &-1.20    &2.50   &-5.38    &0.64\\
N6544    &-1.56    &1.63   &-6.56    &0.73\\
N6541    &-1.83    &2.00   &-8.37    &0.34\\
N6553    &-0.34    &1.17   &-7.99    &0.40\\
N6558    &-1.44    &2.50   &-6.46    &0.00\\
I1276    &-0.73    &1.29   &-6.67    &0.57\\
Ter 12   &-0.50    &0.57   &-4.14    &0.53\\
N6569    &-0.86    &1.27   &-7.88    &0.08\\
N6584    &-1.49    &1.20   &-7.68    &0.85\\
N6624    &-0.42    &2.50   &-7.50    &0.79\\
N6626    &-1.45    &1.67   &-8.33    &0.41\\
N6638    &-0.99    &1.40   &-6.83    &0.20\\
N6637    &-0.71    &1.39   &-7.52    &0.20\\
N6642    &-1.35    &1.99   &-6.57    &0.20\\
N6652    &-0.96    &1.80   &-6.57    &0.38\\
N6656    &-1.64    &1.31   &-8.50    &0.69\\
Pal 8    &-0.48    &1.53   &-5.52    &0.75\\
N6681    &-1.51    &2.50   &-7.11    &0.32\\
N6712    &-1.01    &0.90   &-7.50    &0.54\\
N6715    &-1.59    &1.84  &-10.01    &1.29\\
N6717    &-1.29    &2.07   &-5.67    &0.36\\
N6723    &-1.12    &1.05   &-7.86    &0.41\\
N6749    &-1.60    &0.83   &-6.70    &0.70\\
N6752    &-1.56    &2.50   &-7.73    &0.72\\
N6760    &-0.52    &1.59   &-7.86    &0.68\\
N6779    &-1.94    &1.37   &-7.38    &0.99\\
Ter 7    &-0.58    &1.08   &-5.05    &1.20\\
Pal 10   &-0.10    &0.58   &-5.79    &0.81\\
Arp 2    &-1.76    &0.90   &-5.29    &1.33\\
N6809    &-1.81    &0.76   &-7.55    &0.58\\
Ter 8    &-2.00    &0.60   &-5.05    &1.28\\
Pal 11   &-0.39    &0.69   &-6.81    &0.89\\
N6838    &-0.73    &1.15   &-5.56    &0.83\\
N6864    &-1.32    &1.88   &-8.35    &1.11\\
N6934    &-1.54    &1.53   &-7.65    &1.16\\
N6981    &-1.40    &1.23   &-7.04    &1.11\\
N7006    &-1.63    &1.42   &-7.68    &1.59\\
N7078    &-2.25    &2.50   &-9.17    &1.02\\
N7089    &-1.62    &1.80   &-9.02    &1.02\\
N7099    &-2.12    &2.50   &-7.43    &0.85\\
Pal 12   &-0.94    &1.94   &-4.48    &1.20\\
Pal 13   &-1.65    &0.66   &-3.51    &1.44\\
N7492    &-1.51    &1.00   &-5.77    &1.40\\

\enddata

\end{deluxetable}

\clearpage

\begin{deluxetable}{ccrccrrc}
\tabletypesize{}
\tablecaption{Radial dependence of metallicity\tablenotemark{1}}
\tablewidth{0pt}

\tablehead{&\multicolumn{3}{c}{n} &&\multicolumn{3}{c}{n}\\
\colhead{$\log R$} &\multicolumn{3} {c} {$-2.40<[Fe/H]\leq -1.60$}
&&\multicolumn{3}{c} {$-1.60<[Fe/H]\leq -1.00$}} 

\startdata

 2.00 to  2.19   &&1   &1   &&&1  & 1\\
 1.80 to  1.99   &&2   &2   &&&2  & 2\\
 1.60 to  1.79   &&0   &0   &&&1  &1\\
 1.40 to  1.59   &&6   &6   &&&3  &3\\
 1.20 to  1.39   &&~~~~~~11  &6   &&&~~~~~~~5  &3\\
 1.00 to  1.19   &&5   &5   &&&6  &6\\
 0.80 to  0.99   &&8   &8   &&&6  &6\\
 0.60 to  0.79   &&3   &3   &&&9  &9\\
 0.40 to  0.59   &&5   &5   &&&7  &7\\
 0.20 to  0.39   &&4   &4   &&&6  &6\\
 0.00 to  0.19   &&2   &2   &&&2  &2\\
-0.20 to -0.01   &&0   &0   &&&2  &2\\
-0.40 to -0.21   &&0   &0   &&&2  &2\\

\enddata

\tablenotetext{1}{Numbers in fourth and sixth columns of the table exclude clusters associated with the Sagittarius system.}

\end{deluxetable}

\clearpage

\begin{deluxetable}{rrr}
\tabletypesize{}
\tablecaption{Magnitude dependence of cluster central concentration\label{tbl-3}}
\tablewidth{0pt}
\tablehead{\colhead{$M_V$} & \colhead{$1.50<c<2.00$} & \colhead{$c>2.40$}}
\startdata

-10.00 to -10.99    &1    &0\\
 -9.00 \phm{ to} -9.99    &7    &1\\
 -8.00 \phm{ to} -8.99   &16    &1\\
 -7.00 \phm{ to} -7.99   &15   &12\\
 -6.00 \phm{ to} -6.99    &5    &5\\
 -5.00 \phm{ to} -5.99    &1    &2\\
 -4.00 \phm{ to} -4.99    &1    &0\\
 -3.00 \phm{ to} -3.99    &0    &2\\
 -2.00 \phm{ to} -2.99    &1    &0\\

\enddata

\end{deluxetable}

\end{document}